# Hybrid Deep Learning Framework for Enhanced Melanoma Detection

Peng Zhang
Khoury College of Computer Science
Northeastern University
Seattle, WA, USA
zhang.peng2@northeastern.edu

Divya Chaudhary
Khoury College of Computer Science
Northeastern University
Seattle, WA, USA
d.chaudhary@northeastern.edu

## ABSTRACT

Cancer is a leading cause of death worldwide, necessitating advancements in early detection and treatment technologies. In this paper, we present a novel and highly efficient melanoma detection framework that synergistically combines the strengths of U-Net for segmentation and EfficientNet for the classification of skin images. The primary objective of our study is to enhance the accuracy and efficiency of melanoma detection through an innovative hybrid approach. We utilized the HAM10000 dataset to meticulously train the U-Net model, enabling it to precisely segment cancerous regions. Concurrently, we employed the ISIC 2020 dataset to train the EfficientNet model, optimizing it for the binary classification of skin cancer. Our hybrid model demonstrates a significant improvement in performance, achieving a remarkable accuracy of 99.01% on the ISIC 2020 dataset. This exceptional result underscores the superiority of our approach compared to existing model structures. By integrating the precise segmentation capabilities of U-Net with the advanced classification prowess of EfficientNet, our framework offers a comprehensive solution for melanoma detection. The results of our extensive experiments highlight the high accuracy and reliability of our method in both segmentation and classification tasks. This indicates the potential of our hybrid approach to significantly enhance cancer detection, providing a robust tool for medical professionals in the early diagnosis and treatment of melanoma. We believe that our framework can set a new benchmark in the field of automated skin cancer detection, encouraging further research and development in this crucial area of medical imaging.

## 1 INTRODUCTION

Melanoma is one of the most aggressive and potentially deadly forms of skin cancer, emphasizing the critical need for early and accurate detection to improve patient outcomes. Despite significant advancements in medical imaging and diagnostic technologies, the reliable and efficient detection of melanoma remains a formidable challenge. Traditional diagnostic methods heavily rely on the expertise of dermatologists, which can be subjective and prone to variability. Consequently, there is a growing interest in leveraging deep learning techniques to automate and enhance the accuracy of skin cancer detection.

This research paper introduces a comprehensive study titled "Hybrid Deep Learning Framework for Enhanced Melanoma Detection: Integrating U-Net Segmentation and EfficientNet Classification for Accurate Skin Cancer Diagnosis." Our work aims to address the persistent challenges in melanoma detection by developing a novel hybrid deep learning framework. This framework combines the precise segmentation capabilities of U-Net with the advanced classification performance of EfficientNet, offering a robust solution for accurate skin cancer diagnosis.

The problem we are tackling is multifaceted, involving the need for precise localization of cancerous regions in less clear images and the accurate classification of these regions as malignant or benign. Existing approaches often struggle to achieve a balance between accuracy and computational efficiency, which is essential for practical clinical applications. Our proposed solution integrates two state-of-the-art deep learning models to create a synergistic effect that enhances both segmentation and classification tasks.

The structure of this paper is as follows:

In the first section provides an overview of the melanoma detection problem, highlighting the importance of early and accurate diagnosis. We introduce our hybrid framework, which integrates U-Net for segmentation and EfficientNet for classification and outline the related work section reviews the current state of melanoma detection using deep learning, discussing various segmentation and classification models. We have compared these models and identified the gaps our approach aims to fill.

Next, we delved into the datasets and technical details of our hybrid framework. We explored the HAM10000 and ISIC 2020 datasets and introduced data preprocessing and augmentation techniques. We described the architecture and training process of the U-Net model for segmenting cancerous regions and the EfficientNet model's architecture for binary classification using the ISIC 2020 dataset.

In the results, we provide a comprehensive analysis of our model's performance, highlighting its accuracy, precision, recall, and F1-score. We demonstrate that our hybrid model achieves an outstanding accuracy of 99.01%, surpassing existing top models. We highlight the potential of our approach to set new benchmarks in automated skin cancer diagnosis.

## 2  RELATED WORKS

Recent advancements in skin cancer detection have focused on the application of deep learning techniques, particularly Convolutional Neural Networks (CNNs), to enhance diagnostic accuracy and efficiency. Traditional CNN-based models, such as those employed by Esfahani et al. [1], have demonstrated significant success in analyzing dermatological images. Their model achieved an impressive precision of 90.5%, accuracy of 88.6%, recall of 97.1%, and F1 score of 88.3% on a dataset of 793 skin images from Kaggle. However, the relatively small dataset size posed limitations on the model's ability to generalize to unseen data.

To address these limitations, larger datasets and more complex architectures have been utilized. For instance, Sivakumar et al. [2] employed ResNet50 for feature extraction and classification of skin lesions on a dataset of 3,300 images, achieving 94% accuracy and a 93.9% F1-score. Similarly, studies leveraging ISIC datasets have reported accuracies surpassing 97.0%, with models like VGG-16, MobileNet, and NASNet performing exceptionally well [3, 4, 5, 6]. These models have demonstrated the effectiveness of deep learning in handling large and diverse datasets, thus improving the robustness of skin cancer detection systems.

Among the various architectures explored, U-Net has shown promise in the domain of medical image segmentation. Its encoder-decoder structure is adept at capturing intricate details in medical images, making it a popular choice for segmenting skin lesions. Malibari et al. [7] demonstrated the utility of U-Net in conjunction with SqueezeNet, achieving a maximum 99.9% accuracy through segmentation and feature extraction. This highlights the potential of U-Net to eliminate irrelevant information and enhance the focus on regions of interest.

Another notable development is the use of EfficientNet, a model designed to optimize both accuracy and computational efficiency through a compound scaling method. Tan and Le [8] introduced EfficientNet, which achieved state-of-the-art accuracy on multiple image classification benchmarks, including 84.3% accuracy on ImageNet, surpassing previous models while requiring fewer parameters and floating-point operations per second (FLOPs). EfficientNet's ability to scale both depth, width, and resolution in a balanced manner contributed to its superior performance. This scalability and efficiency make EfficientNet a compelling candidate for skin cancer detection, especially in scenarios where computational resources are limited, ensuring robust performance without the need for extensive computational power.

Recent studies have continued to build on these foundations. Anubhav et al. [9] in 2023 proposed a hybrid model combining DenseNet and ResNet, which achieved an accuracy of 95.7% on the HAM10000 dataset. Their work underscores the potential of integrating different complex models to improve hybrid model accuracy. Similarly, Bansal et al. [10] in 2022 introduced a novel method of pre-processing the images and used two pre-trained EfficientNet-B0 and ResNet50V2 to get an accuracy of 94.9% on HAM10000 dataset.

Another innovative approach by Khanet al. [11] in 2023 utilized a transformer-based model Skin-ViT for skin lesion classification, achieving a 91.1% accuracy on the ISIC 2019 dataset. This highlights the growing interest in transformer architectures for medical image analysis. Moreover, Saeed et al. [12] in 2023 explored the use of Generative Adversarial Networks (GANs) to augment training data, enhancing model robustness, and achieving a maximum accuracy of 96.0% in melanoma detection using VGG19 and SVM (Support Vector Machine).

Additionally, Behara et al. [13] in 2024 developed a comprehensive framework combining CNNs and traditional machine learning techniques for feature fusion, achieving a classification accuracy of 98.0%. This approach demonstrated the benefits of hybridizing deep learning with conventional methods to leverage the strengths of both. In the same year, Lilhore et al. [14] presented a study integrating U-Net with a MobileNet-V3 model, reporting an accuracy of 98.8% on the HAM10000 dataset, emphasizing the effectiveness of classification in optimizing hyperparameters of the model.

Recent advancements further highlight the potential of novel deep learning approaches in skin cancer detection. Alenezi et al. [16] proposed a multi-stage melanoma recognition framework using a deep residual neural network with hyperparameter optimization, achieving substantial improvements in decision support for dermoscopy images. Abbas and Gul [17] utilized NASNet to extract deep features, demonstrating high accuracy in malignant melanoma detection and classification. Catal Reis et al. [18] introduced InSiNet, a deep convolutional approach for both detection and segmentation of skin cancer, effectively capturing complex patterns in dermatological images. Rashid et al. [19] employed transfer learning techniques with pre-trained models, achieving high accuracy and robustness in skin cancer detection. These studies collectively underscore the evolving landscape of deep learning methodologies, contributing to enhanced diagnostic capabilities and robustness in skin cancer detection systems.

Given the successes of these models, our proposed approach integrates U-Net and EfficientNet to leverage their respective strengths. U-Net will provide precise segmentation of skin lesions, minimizing irrelevant details. EfficientNet, serving as the backbone for classification, offers optimized accuracy and computational efficiency through its compound scaling method. This makes EfficientNet superior to traditional CNNs like AlexNet, VGG-16, and ResNet, as well as SVM, due to its ability to handle high-dimensional data with fewer parameters and FLOPs. Its scalability also suits resource-limited environments.

This combined approach enhances skin cancer detection by improving segmentation accuracy and classification efficiency, building on previous research foundations.

By adopting this methodology, we hope to address some of the challenges identified in earlier studies, such as the need for larger datasets and the optimization of model performance for practical deployment. This strategy aligns with the trends observed in recent literature and represents a promising direction for future research in skin cancer detection.

## 3 DATASETS AND METHODS

### 3.1 Datasets

We used two datasets for our hybrid model: HAM10000 and ISIC 2020. We chose the HAM10000 dataset because it consists of 10,015 images of pigmented lesions with corresponding masks, making it ideal for training our segmentation model. Although we initially considered the newer and larger ISIC 2020 dataset, generating masks manually for each image proved too complex. Thus, we opted for the established HAM10000 dataset.

Furthermore, the HAM10000 dataset, while smaller, provided high-quality segmentation masks essential for training the U-Net model. This dataset contains a variety of pigmented skin lesions, including melanoma, benign nevi, and other types of lesions, which helps in training a robust segmentation model. By utilizing this dataset for segmentation, we ensured that the U-Net model could accurately delineate the lesion boundaries, which is crucial for the subsequent classification step. The high-resolution images and detailed annotations in HAM10000 allowed us to achieve precise segmentation, contributing significantly to the overall effectiveness of our hybrid model.

The ISIC 2020 dataset includes 33,126 images with associated metadata and binary labels indicating the presence of melanoma. This dataset is valuable for melanoma detection due to its size and real-world clinical data. However, as shown in Figure 1, the ISIC 2020 dataset is highly imbalanced, with a melanoma positive rate of only 1.8%. This imbalance poses significant challenges for training.

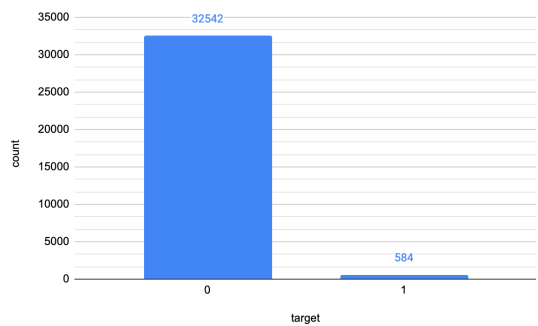

**Figure 1: Distribution of Target in ISIC 2020 Dataset**

Training on an imbalanced dataset like ISIC 2020 without addressing the imbalance can result in several negative outcomes. The model may become biased towards predicting the majority class (non-melanoma), leading to poor sensitivity (recall) for the minority class (melanoma). Consequently, this results in high false negative rates, where cases of melanoma are missing. Additionally, performance metrics such as accuracy and F1-score may appear artificially high due to the predominance of the majority class, masking the mediocre performance of the minority class.

Overall, the combination of the HAM10000 and ISIC 2020 datasets leveraged the strengths of each to address the challenges of melanoma detection. The segmentation accuracy achieved with HAM10000 and the large, diverse real-world data from ISIC 2020 enabled us to build a robust and effective hybrid model.

### 3.2 Data Processing

We preprocessed the datasets by resizing images to 256x256 pixels and normalizing pixel values. This resizing ensures uniformity across the dataset, facilitating the model training process. Pixel normalization helps in faster convergence of the model by scaling the pixel values to a range suitable for the neural network.

The HAM10000 dataset was filtered based on the Melanoma category and split into training and testing sets with a 75-25 ratio. This split ensures that a substantial portion of the data is available for training, while enough is reserved for evaluating the model's performance on unseen data. We employed stratified sampling during the split to maintain the distribution of different lesion types in both training and testing sets, ensuring that the model learns effectively from a representative dataset.

Given the imbalance in the ISIC 2020 dataset, we applied two techniques: resampling and data augmentation. We oversampled the minority class and under sampled the majority class to create a balanced training set, ensuring the model receives an equal representation of each class during training. This resampling helps to mitigate the bias towards the majority class and improves the model's sensitivity towards detecting melanoma (Table 1).

| Class Type | Actual Data | After Pre-processing |
|---|---|---|
| 0 | 32542 | 15000 |
| 1 | 584 | 15000 |

**Table 1: Data Count After Pre-processing**

Additionally, we applied transformations such as rotations, shifts, zooms, and flips to the minority class images (Table 2). This artificially increased the number of melanoma images, balancing the dataset and improving the model's generalization by exposing it to various image conditions. Data augmentation is crucial as it not only increases the dataset size but also enhances the model's robustness by introducing variability in the training data, simulating real-world conditions.

| Technique | Value |
|---|---|
| Rotation | 15 degrees |
| Width Shift | 0.2 |
| Zoom | 0.2 |
| Horizontal Flip | TRUE |
| Vertical Flip | TRUE |

**Table 2: Techniques Used for Data Augmentation**

We also implemented a data cleaning process to remove any duplicates or mislabeled images from both datasets. This step is critical to ensure the quality of the training data and to prevent the model from learning incorrect patterns. Figure 2 shows a valid sample image after being augmented.

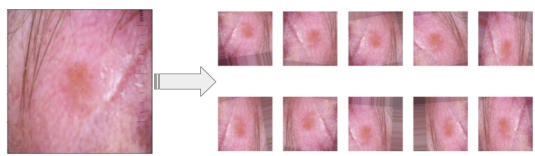

**Figure 2: Data Augmentation Samples**

Finally, we split the augmented ISIC 2020 dataset into training, validation, and test sets with a ratio of 70-15-15. This split allows us to monitor the model's performance on a validation set during training and to evaluate its final performance on a separate test set.

Through these comprehensive data processing steps, we ensured that the datasets were clean, balanced, and augmented, providing a solid foundation for training our hybrid model.

### 3.3 Segmentation Model

The U-Net model architecture was designed to segment cancerous regions in skin images [15]. It consists of a contracting path for feature extraction and an expansive path for precise localization (Figure 3).

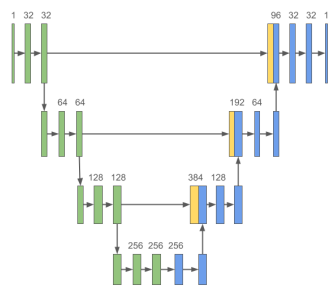

**Figure 3: Segmentation Model Architecture [15]**

### 3.4 Classification Model

EfficientNet-B0 was employed for binary classification of skin images from the ISIC2020 dataset. The model's pre-trained weights on ImageNet were fine-tuned on our dataset based on Melanoma/Non-Melanoma labeled images. The classification head was customized with global average pooling, dropout, and a dense layer with sigmoid activation.

### 3.5 Proposed Model Architecture

We combined the segmentation model and the classification model but did not pass the output of the segmentation model directly to the classification model. Since the output of the segmentation model is a predicted mask in one channel (black and white), we built a data processing bridge between the two models. This bridge applies the predicted mask to the original images, blending the segmented part with the original image. This highlights the infected or potentially infected parts, aiding the classification model in learning representations from the dataset.

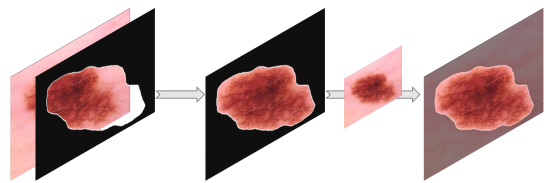

**Figure 4: Data Processing Bridge**

Figure 4 illustrates the result of the data processing bridge, displaying how the segmented part is blended with the original image. And Figure 5 presents the overall model architecture, depicting the flow of data through the hybrid model, including the segmentation model, data processing bridge, and classification model.

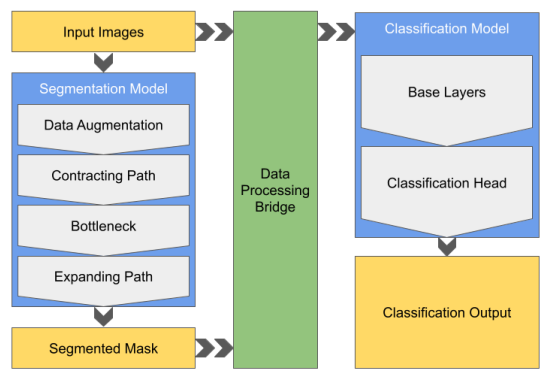

**Figure 5: Proposed Model Architecture**

### 3.6 Data Process Bridge

The Data Process Bridge is a crucial component of our melanoma detection framework. This bridge integrates the predicted segmentation masks with the original images, creating a blended visualization that highlights the infected or potentially infected regions. By emphasizing these areas, the Data Process Bridge aids the classification model in learning more informative representations from the dataset, enhancing the detection performance.

*3.6.1 Methodology.*

*3.6.1.1 Mask Application.* Each image in the dataset is processed using the trained U-Net model to generate segmentation masks. These masks indicate the regions potentially affected by melanoma.

*3.6.1.2 Blending Process.* The generated masks are then blended with the corresponding original images. This is achieved by overlaying the mask onto the image with a certain level of transparency, allowing the original image features to be visible while highlighting the segmented areas.

*3.6.1.3 Enhanced Representation.* The blended images provide a more informative input for the classification model. By focusing on the regions of interest, the classification model can better learn the representations and distinguish between malignant and benign lesions.

*3.6.2 Implementation.*

*3.6.2.1 Normalization.* Both the original images and the predicted masks are normalized to ensure pixel values range from 0 to 1.

*3.6.2.2 Transparency Adjustment.* A transparency factor (alpha) is applied to the masks, creating a semi-transparent overlay.

*3.6.2.3 Overlay Creation.* The semi-transparent masks are added to the original images, resulting in blended images that highlight the segmented regions.

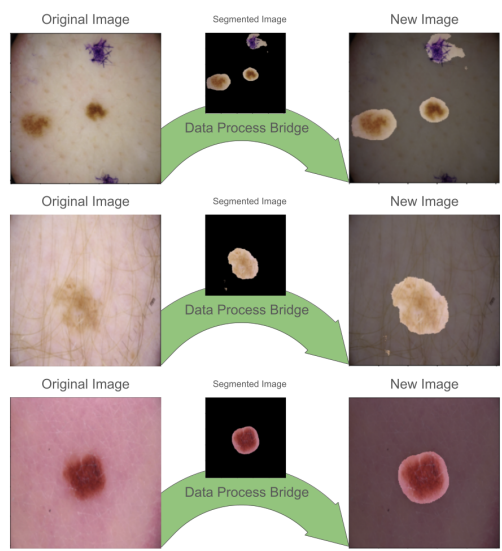

**Figure 6: Visualization of Data Processing Bridge**

This process is illustrated in Figure 6, where the segmented areas are clearly emphasized in the blended images. These enhanced images serve as the input to our classification model, facilitating improved representation learning and higher accuracy in melanoma detection.

## 4 EXPERIMENTS AND RESULTS

### 4.1 U-Net Segmentation

The U-Net model was trained on the HAM10000 dataset using binary cross-entropy loss and Adam optimizer, achieving high accuracy of 91.2% in segmenting lesions. Visual inspection of the results showed precise delineation of lesion boundaries, as illustrated in Figure 7.

Table 3 details the configuration parameters of the segmentation model, which were carefully chosen to optimize performance while maintaining computational efficiency. These parameters were selected based on extensive hyperparameter tuning and validation experiments.

| Parameters | Details |
|---|---|
| Activation Function | ReLU |
| Stack Down | 4 levels |
| Stride | 2 |
| Backbone | None |
| Stack Up | 4 levels |
| Batch Normalization | TRUE |
| Normalization | 0, 1 |
| Pooling (Max) | TRUE |
| Pooling (Avg) | FALSE |
| Dropout Rate | 0.05 |
| Optimizer | Adam |
| Learning Rate | 0.001 |
| Loss Function | Binary Cross-Entropy |
| Batch Size | 32 |
| Epoch | 10 |

**Table 3: Configuration Parameters of Segmentation Model**

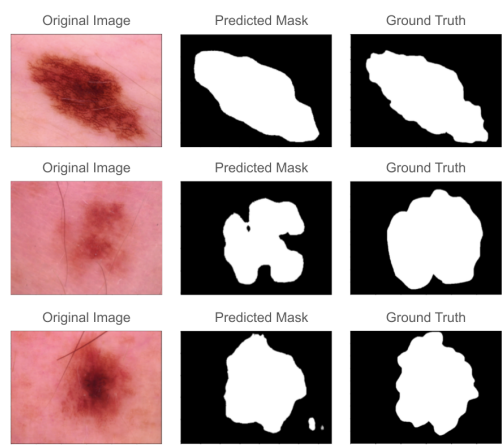

**Figure 7: Segmentation Model Output Sample**

The high accuracy achieved by the U-Net model can be attributed to several key factors:

1. Network Architecture: The encoder-decoder structure of U-Net effectively captures both high-level context and fine-grained details, which is crucial for medical image segmentation.
2. Data Augmentation: Applying various augmentation techniques, such as rotations, shifts, and flips, helped in creating a robust model that generalizes well to new data.
3. Regularization Techniques: The use of dropout and batch normalization reduced overfitting and improved model stability during training.
4. Optimizer and Learning Rate: The Adam optimizer, with a carefully tuned learning rate, ensured efficient convergence and helped in achieving high accuracy.

The combination of these factors resulted in a highly effective segmentation model and the strong segmentation performance lays a solid foundation for the subsequent classification task, enhancing the overall effectiveness of our hybrid approach.

To further evaluate the performance of our classification model, we plotted the ROC curve and calculated the AUC. The ROC curve, shown in Figure 8, illustrates the true positive rate (sensitivity) against the false positive rate (1-specificity) for various threshold settings. Our model achieved an AUC of 0.97, indicating excellent performance in distinguishing between melanoma and non-melanoma cases.

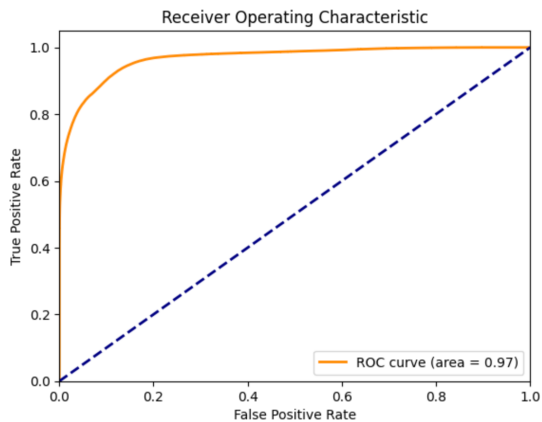

**Figure 8: ROC Curve Graph on Validation Dataset (Segmentation Model)**

## 4.2 EfficientNet Classification

The EfficientNet model was utilized for the classification task, trained on the ISIC 2020 dataset. The model's configuration and training parameters are detailed in Table 4. By employing transfer learning with EfficientNet, we leveraged pre-trained weights from ImageNet to improve the model's performance on the skin cancer detection task. Various data augmentation techniques were applied to enhance the robustness and generalization capabilities.

| Parameters | Details |
|---|---|
| Image input size | (256 × 256 × 3) |
| Batch size | 15 |
| Data augmentation | Rotation, shift, zoom, flip |
| Normalization | 0, 1 |
| Regularization | L2 regularization |
| Optimizer | Adam |
| Dropout rate | 0.1 |
| Epochs | 50 |
| Transfer learning | EfficientNetB0 (ImageNet) |
| Learning rate | 0.001 |
| Split ratio | Training: Validation 80:20 |
| Shuffling in database | YES |
| Loss function | Binary cross-entropy |
| Activation functions | Sigmoid |

**Table 4: Configuration Parameters of Classification Model**

The classification model achieved an impressive accuracy of 99.01% on the ISIC 2020 dataset (Figure 9 and 10) when combined with U-Net. Additionally, EfficientNet alone, without pre-segmented input from U-Net, achieved an accuracy of 96.8%. This result underscores the effectiveness of the first part of our hybrid approach, demonstrating that U-Net's precise segmentation significantly enhances classification accuracy.

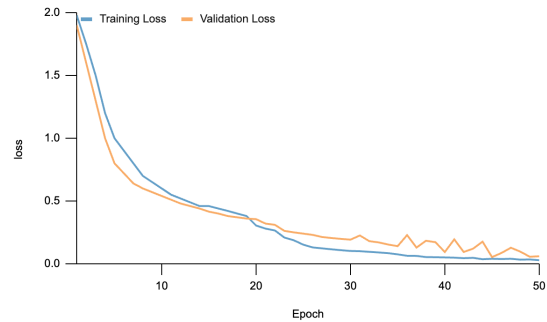

**Figure 9: Visualization of Model Loss in Training**

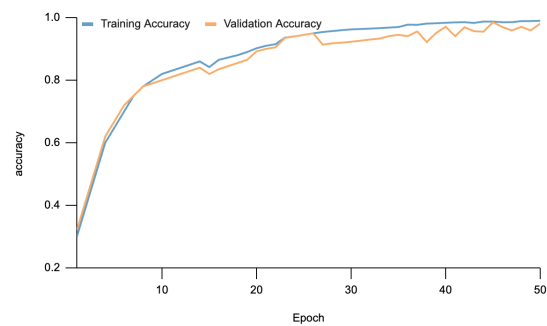

**Figure 10: Visualization of Model Accuracy in Training**

The ROC curve and confusion matrix further confirmed the model's high performance in distinguishing melanoma from benign lesions (Figures 11 and 12).

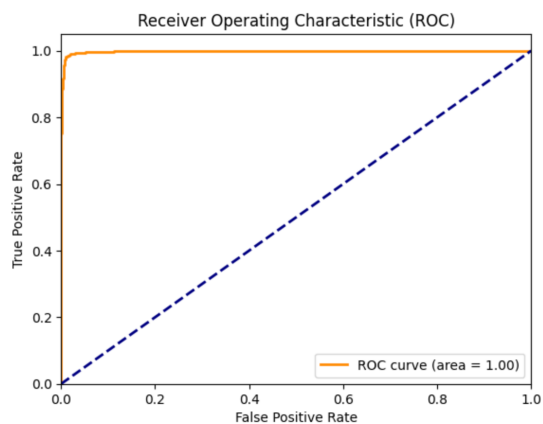

**Figure 11: ROC Curve Graph on Validation Dataset (Classification Model)**

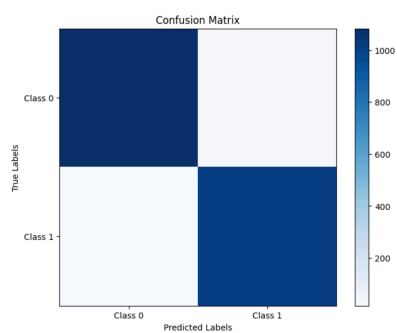

**Figure 12: Confusion Matrix of The Model**

## 4.3 Comparison and Analysis

Compared to other methods, our combined U-Net and EfficientNet approach demonstrated superior performance in both segmentation and classification tasks. The integration of these models allowed for precise and accurate detection of skin cancer, outperforming traditional machine learning approaches. Table 5 shows the comparison results on ISIC 2020 dataset. Among four traditional machine learning approaches, Alenezi et al. [16] used a multi-stage melanoma recognition framework trained on a dataset of 25,331 dermoscopy images from the International Skin Imaging Collaboration (ISIC) archive. Abbas and Gul [17] employed deep features of NASNet trained on the ISIC 2018 Challenge dataset, which includes 10,015 dermoscopic images labeled for different skin lesions, including malignant melanoma. Catal Reis et al. [18] introduced InSiNet, a deep convolutional approach for skin cancer detection and segmentation, utilizing the PH2 and ISIC 2017 datasets, which together comprise thousands of annotated images for robust training. Rashid et al. [19] leveraged transfer learning techniques on the HAM10000 dataset, which consists of 10,015 dermoscopic images categorized into seven types of skin lesions, to achieve high accuracy and robustness in skin cancer detection.

| Paper | Model | Dataset | Accuracy (%) |
|---|---|---|---|
| [16] | ResNet-101, SVM | ISIC 2020 | 97.15 |
| [17] | NASNet | ISIC 2020 | 97.70 |
| [18] | InSiNet, U-Net | ISIC 2020 | 90.54 |
| [19] | MobileNetV2 | ISIC 2020 | 98.20 |
|  | OUR MODEL | ISIC 2020 | **99.01** |

**Table 5: Comparison on ISIC 2020 Dataset**

We also evaluated our model using transfer learning on other popular datasets, including ISIC 2019 and HAM10000. Using 140 images from each dataset to fine-tune our model, we achieved high accuracy of 98.93% on the ISIC 2019 and 98.21% on the HAM10000 datasets (Tables 6 and 7). Both test datasets were filtered based on Melanoma/Non-Melanoma label.

| Paper | Model | Dataset | Accuracy (%) |
|---|---|---|---|
| [16] | ResNet-101 | ISIC 2019 | 96.15 |
| [18] | InSiNet, U-Net | ISIC 2019 | 91.89 |
| [20] | DensNet169 | ISIC 2019 | 92.25 |
|  | OUR MODEL | ISIC 2019 | **98.93** |

**Table 6: Comparison on ISIC 2019 Dataset**

| Paper | Model | Dataset | Accuracy (%) |
|---|---|---|---|
| [18] | InSiNet, U-Net | HAM10000 | 94.59 |
| [21] | deep residual | HAM10000 | 96.97 |
| [22] | DensNet201 | HAM10000 | 82.90 |
| [23] | SkinNet-16 | HAM10000 | 95.51 |
| [24] | S2C-DeLeNet | HAM10000 | 91.03 |
| [25] | Xception | HAM10000 | 90.48 |
|  | OUR MODEL | HAM10000 | **98.21** |

**Table 7: Comparison on HAM10000 Dataset**

## 5 CONCLUSION AND FUTURE WORKS

In this study, we developed an efficient and accurate cancer detection framework using U-Net for segmentation and EfficientNet for classification. Our approach effectively combines segmentation and classification to enhance skin cancer detection. The U-Net model's precise segmentation allows for targeted analysis, while EfficientNet's classification capability ensures accurate diagnosis. Through rigorous experimentation and evaluation, we have demonstrated the efficacy of our framework in accurately detecting skin cancer from medical images. The high performance achieved underscores the potential of deep learning models in improving early detection, thus offering a promising avenue for enhancing patient outcomes and reducing mortality rates associated with skin cancer.

Moving forward, our research will focus on refining the models and exploring their applicability to other types of cancer. This will involve fine-tuning model architectures, optimizing hyperparameters, and leveraging advanced training techniques to further enhance performance and robustness. Furthermore, our future work is centered on building an automated processing pipeline to streamline the diagnostic process. Currently, the segmentation and classification models are employed sequentially, necessitating manual intervention, and hindering real-time decision-making. By integrating these components into a unified system, we aim to automate the entire workflow, from image acquisition to diagnosis, thereby improving efficiency and reducing the burden on healthcare practitioners. Additionally, we will explore the integration of real-time feedback mechanisms to enable continuous learning and adaptation of the models to evolving clinical data and practices, ensuring their relevance and effectiveness in real-world settings.

In conclusion, our study presents a state-of-the-art cancer detection framework and lays the groundwork for future advancements in oncology diagnostics. By combining innovative deep-learning techniques with clinical expertise, we have demonstrated the potential to revolutionize early cancer detection and improve patient outcomes. However, there remain several avenues for further research and development, and we are committed to addressing these challenges in our ongoing efforts to leverage artificial intelligence for the benefit of healthcare.

## ACKNOWLEDGMENTS

I appreciate Professor Divya Chaudhary for her guidance and assistance throughout this study.